# A Novel end-to-end Digital Health System Using Deep Learning-based ECG Analysis

Artemis Kontou[1], Natalia Miroshnikova[2], Costakis Matheou[2], Sophocles Sophocleous[2], Nicholas Tsekouras[2], Kleanthis Malialis[1], Panayiotis Kolios[1,3]

[1] KIOS Research and Innovation Center of Excellence (KIOS CoE), University of Cyprus, Panepistimiou 1, Aglantzia 2109, Cyprus
[2] MEDTL Medical Technologies Ltd., 91 Limassol Avenue, Aglantzia, Office 202, Nicosia 2121, Cyprus
[3] Department of Computer Science, University of Cyprus, Panepistimiou 1, Aglantzia 2109, Cyprus

Corresponding author: A. Kontou (e-mail: kontou.artemis@ucy.ac.cy).

**ABSTRACT**

This study presents AI-HEART, a cloud-based information system for managing and analysing long-duration ambulatory electrocardiogram (ECG) recordings and supporting clinician decision-making. The platform operationalises an end-to-end pipeline that ingests multi-day three-lead ECGs, normalises inputs, performs signal preprocessing, and applies dedicated deep neural networks for wave delineation, noise/quality detection, and beat- and rhythm-level multi-class arrhythmia classification. To address class imbalance and real-world signal variability, model development combines large clinically annotated datasets with expert-in-the-loop curation and generative augmentation for under-represented rhythms. Empirical evaluation on three-lead ambulatory ECG data shows that delineation accuracy is sufficient for automated interval measurement, noise detection reliably flags poor-quality segments, and arrhythmia classification achieves high specificity with clinically useful macro-averaged performance across common and rarer rhythms. Beyond predictive accuracy, AI-HEART provides a scalable deployment approach for integrating AI into routine ECG services, enabling traceable outputs, audit-friendly storage of recordings and derived annotations, and clinician review/editing that captures feedback for controlled model improvement. The findings demonstrate the technical feasibility and operational value of a noise-aware AI-ECG platform as a digital health information system.

**Acknowledgement:** This work has been funded by the Research and Innovation Foundation (RIF) of Cyprus under the RESTART 2016–2020 Programmes for Research, Technological Development and Innovation, through the «DISRUPT» Programme, Call DISRUPT/0123(C), Project AI-HEART (Project No. DISRUPT/0123(C)/0014), which is co-financed by the Recovery and Resilience Facility of the European Union and the Republic of Cyprus.

## 1. Introduction

Cardiovascular diseases (CVDs) remain the leading cause of mortality globally, accounting for an estimated 18.6 million deaths annually [1]. This burden disproportionately impacts health systems in low- and middle- income countries, where access to specialist cardiology services is often constrained [2]. The expanding use of ambulatory electrocardiogram (ECG) monitoring further increases diagnostic workload, making timely and standardised review of large ECG datasets a practical necessity [3]. In response, deep learning has rapidly advanced automated ECG interpretation, enabling scalable detection of arrhythmias and conduction abnormalities from large volumes of real-world recordings. However, deployment in routine clinical workflows remains constrained by variable signal quality, noise and artefacts, and the need for robust performance across heterogeneous devices and recording conditions [4]. These challenges are particularly acute in ambulatory monitoring, where recordings are typically acquired continuously using two or three leads over extended periods, and clinically relevant events may be intermittent and easily missed in brief snapshots. Developing noise-robust deep learning models tailored to three-lead ambulatory ECG therefore addresses a practical bottleneck for real-world screening and monitoring: accurate multi-class detection under non-ideal conditions, with outputs that can support timely clinical review and prioritisation.

Early rule-based algorithms and shallow machine-learning methods have been largely superseded by convolutional neural networks (CNNs) and recurrent neural networks (RNNs), which learn hierarchical features directly from raw or minimally processed ECG data [5]. For example, Hannun et al. demonstrated cardiologist-level multi-class arrhythmia classification from single-lead ambulatory ECG recordings using an end-to-end deep neural network, while Ribeiro et al. showed that residual convolutional networks (ResNet-style CNNs) can achieve high-accuracy multi-label classification from 12-lead ECGs at scale [5, 9]. Seminal work by Jin et al. [7] introduced an interpretable, knowledge-fused deep neural network for single-lead (lead II) ECG that achieved performance comparable to or exceeding cardiologists for multi-class arrhythmia diagnosis, and subsequent studies have adapted architectures such as ResNet and DenseNet for multi-lead ECG classification, achieving state-of-the-art results on public datasets including PTB-XL [8, 9]. Beyond rhythm disorders, deep learning models have shown remarkable capability in extracting "hidden" biomarkers from standard ECGs. Notably, an AI-ECG algorithm for detecting asymptomatic low ejection fraction (LEF) achieved an area under the receiver operating characteristic curve (AUC) of ~0.93 and has received U.S. FDA 510(k) clearance, paving the way for regulatory approval of AI-derived diagnostic software [10, 11].

Translational progress is evidenced by the emergence of commercial and near-commercial AI-ECG platforms integrated into clinical workflows. These include cloud-base Software as a Medical Devices (SaMD) for ambulatory recordings analysis such as DeepRhythmAI, Cardiologs and BeatLogic that provide automated, high-accuracy reporting for major arrhythmias in real-world settings [6, 12, 13]. Such platforms have moved beyond proof-of-concept to demonstrate practical utility in hospital-based decision support. Regulatory bodies, including the FDA and European CE marking authorities, have established pathways for SaMD, with several AI-ECG algorithms now cleared for clinical use, underscoring the field's maturation [14].

Despite the remarkable progress in algorithmic performance and regulatory advancement, the translation of AI-ECG research into robust, generalizable, and clinically versatile tools faces several intertwined and persistent limitations. A significant portion of commercial and research efforts remains narrowly optimized for single, high-prevalence conditions such as AF [15, 16]. While effective for targeted screening, this focused development leaves a substantial diagnostic gap for the broader spectrum of arrhythmias and conduction disorders including complex ectopy, supraventricular tachycardia, and various degrees of atrioventricular block that are essential for comprehensive ECG interpretation in general practice. Models trained on limited rhythm sets often fail to generalize to the wider differential diagnoses required in routine clinical workflows, limiting their utility as all-purpose decision-support tools.

Furthermore, the challenge of real-world signal quality is frequently under-addressed. Many reported high-performing ECG AI models are developed and validated on relatively clean or curated datasets where noise is commonly handled indirectly (e.g., quality filtering/exclusion) rather than being modeled as a first-class part of the pipeline, which can limit real-world performance [4, 17]. This oversight is critical, as motion artifacts, baseline wander, and electrode contact issues are ubiquitous in ambulatory, emergency, and primary care settings. The lack of integrated, learnable noise-detection modules leaves systems vulnerable to performance degradation and erroneous interpretations when faced with poor-quality recordings, undermining their reliability outside controlled environments [18].

Equally concerning is the static nature of many AI-ECG solutions, which are often developed as fixed products without built-in mechanisms for continuous learning and lifecycle management [19]. Post-deployment, models can suffer from dataset shift, struggle with rare but clinically critical arrhythmias due to class imbalance, and lack pathways to incorporate new expert knowledge. Few platforms describe a regulatory-grade framework that supports systematic dataset curation, expert-in-the-loop feedback, and version-controlled model updates capabilities that are essential for maintaining safety and efficacy over time and aligning with evolving regulatory expectations for adaptive AI.

Additionally, many existing solutions are tightly coupled to specific acquisition hardware, proprietary data formats, or high-resource healthcare IT ecosystems [20]. This hardware and software dependence creates interoperability barriers, limits scalability across diverse clinical settings particularly in low-resource environments and hinders integration into heterogeneous hospital information systems. The vision of broadly accessible AI-assisted ECG analysis requires platform-agnostic, cloud-based architectures that can ingest standardized data from diverse sources and scale efficiently across different healthcare infrastructures [21]. These collective gaps narrow scope, noise sensitivity, static design, and ecosystem dependence currently hinder the development of AI-ECG systems that are truly fit for widespread, sustainable, and trusted integration into global healthcare delivery.

This study is designed to address these limitations directly. It proposes a cloud-based SaMD platform called AI-Heart that integrates a suite of deep learning models for delineation, noise detection, and multi-class arrhythmia classification into a cohesive, end-to-end workflow. Its architecture explicitly prioritizes broad rhythm coverage, explicit noise robustness through dedicated noise-handling components, and a regulatory-grade framework supporting continuous learning via generative data augmentation and expert-in-the-loop curation tools. By adopting a hardware-agnostic, scalable cloud deployment model, the platform aims to be suitable for diverse clinical settings, from hospitals to underserved regions.

The objective of this paper is twofold. First, we examine the technical feasibility and internal performance of the platform as a cloud-based system that integrates distinct deep-learning models for ECG delineation, noise detection, beat-level classification and rhythm-level arrhythmia classification into a single, end-to-end workflow. Second, we analyse how this architecture functions as a digital health information system, focusing on its ability to ingest and manage long-duration three-lead ambulatory ECG data, support human-in-the-loop decision-making and interface with existing clinical reporting processes. We describe the system architecture and processing pipeline, report quantitative performance for each model component on internal datasets and situate the platform within current AI-ECG and digital health literature, with particular emphasis on noise robustness, comprehensive rhythm coverage and deployment as a scalable, cloud-hosted decision-support service. The remainder of this paper is structured as follows: Section II details the materials and methods, Section III describes the system architecture, Section IV presents experimental results, Section V discusses implications and limitations, and Section VI concludes.

## 2. Data and Methods

### 2.1. System implementation

AI-HEART is implemented as a cloud-based software-as-a-medical-device (SaMD) platform that operationalises a multi-stage ECG analysis pipeline within a secure web application. In practice, the platform combines (i) a browser-based clinician interface for uploading recordings and reviewing outputs, (ii) an application service that exposes REST APIs and orchestrates validation, job execution, persistence and audit logging, and (iii) a containerised AI inference service that executes preprocessing and deep-learning model inference in an isolated runtime (GPU-enabled when available). ECG files and derived artifacts are managed via an object-storage layer, while structured entities, metadata, results and clinician edits are stored in a relational database; authentication and role-based authorisation are enforced via an identity and access management service, fronted by a reverse proxy for secure traffic routing and TLS termination. This implementation approach supports modular updates of model components under configuration control and enables scalable batch processing, while the detailed architecture, component responsibilities and end-to-end data flow are described in Section 3.

### 2.2. Datasets and annotation

All data used for model development and testing were de-identified prior to processing and handled under appropriate governance procedures. ECG recordings originated from routine ambulatory monitoring workflows from more than 10,000 patients and were acquired using multi-day ambulatory monitors. Recordings consisted of long-duration three-lead electrocardiograms, typically spanning many hours to several days per patient and covering a wide range of heart rates, activity levels and signal-quality conditions.

The primary training resource is a proprietary MEDTL repository containing more than 1.5 million manually annotated beats drawn from long-term ECG recordings and covering a target set of 35 beat and rhythm types, including sinus and junctional rhythms, supraventricular and ventricular ectopy, atrial fibrillation and flutter, supraventricular tachycardia, multiple degrees of atrioventricular block and conduction abnormalities. This in-house dataset is complemented by public benchmark databases, including MIT-BIH [23] and AHA [24], which were incorporated to increase population diversity and signal variability and to reduce overfitting to any single device or acquisition protocol.

Reference annotations were generated by trained clinicians following guidelines aligned with conventional ECG interpretation standards. For delineation, annotators marked the onset and offset of P waves, QRS complexes and T waves on selected leads. For noise detection, short segments were labelled as "clean" or "noisy" according to their suitability for diagnostic interpretation. For beat-level classification, each beat received a morphological label from the 35-type taxonomy, while rhythm-level labels were assigned to contiguous 10-second segments, capturing patterns such as atrial fibrillation, atrial flutter, AV blocks, supraventricular runs and ventricular tachycardia episodes. A subset of records underwent double reading, and disagreements were resolved by consensus to reduce label noise.

For the experiments reported in this paper, we focus on two labeled ECG subsets derived from the full taxonomy. The beat classification subset consists of beat-centered annotated ECG segments with label assigned to this beat. Rhythm classification subset consists of 10 seconds frames capturing arrhythmia episodes. Consequently, some rhythms such as atrial fibrillation (AF) and atrial flutter (AFL) appear in both tasks: at the beat level, labels are assigned to individual beats occurring within arrhythmic episodes, whereas at the rhythm level the model classifies 10-second windows as AF, AFL, and so on. These subsets were chosen to balance clinical relevance with sufficient sample sizes, while still including rare but important conduction disturbances.

The dataset was partitioned into training, validation and test sets at the patient level, so that no ECG from a given patient appears in more than one split. An 80/10/10 split was used for training, validation and testing, respectively. The validation set was used for hyperparameter tuning and early stopping, whereas the held-out test set was reserved for the final performance evaluation reported in Section IV.

### 2.3. Signal Pre-processing and Delineation

All ECG recordings undergo standardised preprocessing prior to model inference. A band-pass digital filter is applied to reduce baseline wander, high-frequency noise and power-line interference, and signals are resampled or segmented as needed to match the input requirements of the downstream models.

For wave delineation, AI-HEART employs a deep-learning model that operates on fixed-length windows of multi-lead ECG data and predicts the position of isoelectric segments and P, QRS and T waves. The model maps raw or lightly processed samples to a sequence of labels at the sample or small-frame level, and its outputs are post-processed to derive onset and offset times for each wave. From these fiducial points, clinically relevant intervals such as PR, QRS and QT can be computed (Figure 1).

Delineation quality is assessed by comparing predicted boundary locations with reference annotations on the internal test set. Subsequently we report per-sample accuracy, tolerant accuracy within a small temporal window, F1-score and area under the receiver operating characteristic curve (AUC), demonstrating performance adequate for automated interval measurement and downstream rhythm classification.

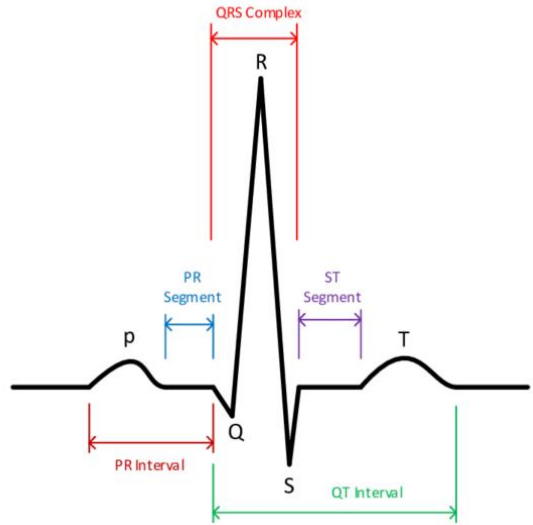

Fig. 1. Fiducial points and main intervals of a typical heartbeat.

### 2.4. Noise Detection

To reduce the impact of artefacts on downstream analysis, AI-HEART includes a dedicated noise-detection model that classifies short ECG segments according to their suitability for diagnostic interpretation. The model is applied in a sliding-window fashion across each recording, and segments identified as low quality are either excluded from automated arrhythmia analysis or clearly flagged to the clinician.

### 2.5. Arrhythmia and Beats Classification

The arrhythmia classifier is designed to recognise a broad set of ECG rhythm classes, including both common and relatively rare patterns. In the current implementation, the model operates on 10-second ECG segments derived from a single diagnostic lead (lead II). These segments are preprocessed and quality-screened by the upstream denoising, delineation and signal-quality modules, then z-normalised before being fed into the classifier. The same architecture is used for both beat-level and rhythm-level classification tasks, with different hyperparameters, enabling consistent treatment of local morphology and longer rhythm episodes.

The classifier employs a hybrid deep-learning architecture that combines convolutional neural networks (CNNs) with Transformer encoders. The CNN blocks act as feature extractors, learning local morphological patterns such as QRS shape, P-wave presence and ST–T abnormalities, and providing robustness to small temporal shifts and baseline variations. The Transformer encoder then applies multi-head self-attention and positional encoding to model long-range temporal dependencies across the 10-second window, allowing the network to integrate beat-to-beat information and capture rhythm-level characteristics such as irregular RR intervals or runs of ectopy. The high-level architecture consists of stacked convolutional layers, followed by a Transformer encoder and one or more fully connected layers with a final softmax output. A high-level diagram of the hybrid CNN–Transformer architecture used for beat-level and rhythm-level arrhythmia classification is shown in Fig. 2, where convolutional blocks extract local morphological features and the Transformer encoder models long-range temporal dependencies across the 10-second ECG window.

The output layer produces a probability distribution over the predefined set of rhythm classes used in this study: atrial fibrillation (AF), atrial flutter (AFL), normal sinus beats (N), junctional beats (J), supraventricular ectopic beats (S), ventricular ectopic beats (V), sinus arrhythmia (SA), supraventricular tachycardia (SVT), several types of atrioventricular block (AV1, AV21, AV22, AV3), aberrancy beats (A) and an unspecified/noise class (X).

During training, the models are using categorical cross entropy loss function with focal loss. Focal loss is used to increase the impact of rare arrhythmias and to emphasise harder examples. Optimiser is AdamW. Training is performed on large annotated datasets comprising both internal MEDTL data and public benchmark databases as emphasized above, with standard regularisation techniques such as dropout a applied to reduce overfitting.

During inference, the classifier generates class probabilities for each analysed 10-second window. These probabilities are then aggregated across the full recording to derive a dominant rhythm and to identify additional arrhythmia episodes, such as paroxysmal AF, runs of SVT or periods of high ventricular ectopic burden. An online platform with visual interfaces presents these results to the clinical user as machine-generated interpretations, with associated confidence scores, which clinicians can review, confirm, modify or override before finalising the report.

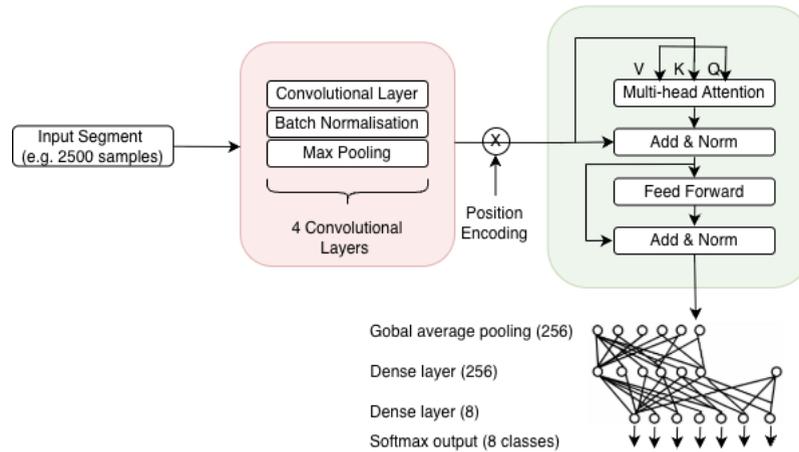

Fig. 2. Hybrid CNN–Transformer architecture used for beat-level and rhythm-level arrhythmia classification in AI-HEART.

*2.6. Data Augmentation and Dataset Curation*

To improve performance, particularly for under-represented arrhythmia classes, AI-HEART employs data augmentation strategies based on generative adversarial networks (GANs). Synthetic ECG beats or segments are generated for rare classes using GANs trained to approximate the distribution of real signals and are then incorporated into the training set under controlled conditions. Internal experiments have shown that this approach can increase classification accuracy and sensitivity for rare rhythms without degrading performance on common classes.

In addition, a custom "Dataset Tornado" application has been developed to support explainability and dataset optimisation. This tool computes low-dimensional embeddings of ECG segments using manifold-learning techniques and visualises them in an interactive interface. Users can explore clusters corresponding to different arrhythmia classes, identify regions with low-confidence or inconsistent predictions and correct mislabelled examples. Corrected labels and curated subsets are then fed back into the training pipeline as part of a continuous learning loop.

*2.7. Evaluation Metrics*

For evaluation purposes, standard classification metrics are computed at both class and aggregate levels. For the delineation model, per-sample accuracy, tolerant accuracy, F1-score and AUC are reported. For the noise-detection model, accuracy, precision, recall, F1-score and AUC are calculated, and confusion matrices were used to illustrate error patterns.

For arrhythmia classification, per-class sensitivity (recall), specificity, precision, F1-score and overall accuracy are reported for both the beat and rhythm sets. Macro-averaged and micro-averaged metrics are computed to summarise performance across classes with different prevalences. Where relevant, models trained with and without GAN-based augmentation and curated labels are compared to quantify the impact of these techniques on under-represented classes. We report 95% confidence intervals (CIs) for proportion-based metrics (sensitivity, specificity, precision, accuracy) using Wilson score intervals. For F1-score, CIs were estimated via non-parametric bootstrap resampling.

## 3. System Architecture and Software Platform

AI-HEART is a proprietary system developed by MEDTL Medical Technologies. For intellectual property and regulatory reasons, only a high-level description of the model architecture and implementation is provided here.

*3.1. Overall Architecture*

Figure 3 summarises the AI-HEART reference architecture and the separation of concerns between (i) the clinician-facing web client, (ii) the application layer responsible for request handling, validation, persistence and orchestration, and (iii) the AI services that execute preprocessing and model inference. Data are split between object storage for ECG files and derived artifacts and a relational datastore for users, metadata, results and audit trails; identity and access management is enforced at the platform boundary to support role-based access control. ECG recordings and associated metadata are received from external acquisition devices or intermediate systems and uploaded to the platform through secure endpoints. The application layer validates and normalises inputs, stores the recordings in encrypted form and schedules AI analysis jobs. The AI services are encapsulated in dedicated components that can be scaled independently, allowing the platform to handle varying workloads without affecting the user-facing interface.

The system is designed to support deployment in multiple geographic regions, so that data can be hosted close to the point of care and in compliance with local data-residency requirements. Internal routing mechanisms ensure that requests are directed to the appropriate regional instance while maintaining a consistent user experience.

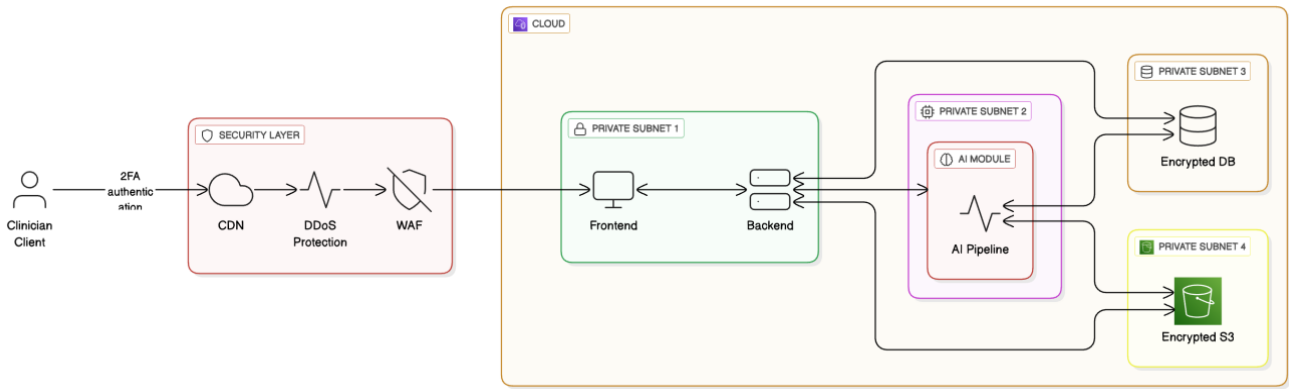

Fig. 3. System architecture of the AI-HEART cloud platform.

### 3.2. Application and AI Services

The application layer implements the platform's external interface that mediates all interactions between the user interface, the data store and the AI components. It implements business logic such as user and organisation management, case creation, credit and billing operations, and orchestration of analysis workflows. By centralising these responsibilities in a well-defined API, client applications can remain thin and independent of internal implementation details.

The AI layer encapsulates the deep-learning models described in Section 2. Model inference runs in a separate service that receives ECG data from the application layer, performs preprocessing and calls the delineation, noise-detection and arrhythmia-classification models. Results are returned as structured annotations and summary findings, which the application layer persists and exposes to the user interface. This separation allows models to be updated, versioned and rolled back under configuration control without modifying the rest of the platform.

Background jobs handle tasks such as batch re-analysis with updated models, generation of summary statistics and periodic health checks. All such jobs are logged and auditable, in line with the quality-management requirements for medical-device software.

### 3.3. Data Management, Security and Access Control

The platform maintains a clear separation between clinical data, operational data and logs. ECG recordings and patient-related metadata are stored in secure, access-controlled repositories with encryption at rest and in transit. Application data such as user accounts, organisations, case status and billing information is maintained in a transactional database, while technical logs and audit trails are stored in dedicated logging systems. These design priorities align with hospital management perspectives that emphasise data quality management, metadata management, and data security as core domains for effective healthcare data governance [22].

User authentication relies on industry-standard mechanisms, including strong password policies and multi-factor authentication. Authorisation is implemented through role-based access control, with distinct roles for system administrators, organisational managers, clinical users and trial coordinators. Each role is associated with a defined set of permissions controlling access to patients, recordings and configuration options.

Every clinically relevant action, such as viewing a case, editing a report or exporting data, is recorded in an audit log with user identity, timestamp and action details. These logs support traceability, incident investigation and regulatory compliance.

### 3.4. Clinical Workflow And User Interface

AI-HEART is integrated into the clinical workflow as a decision-support tool rather than a fully automated reader. When a new ambulatory ECG recording is uploaded and associated with a patient and requesting clinician, the system performs automated analysis in the background, including preprocessing, delineation, noise handling and multi-class arrhythmia classification. Once processing is complete, the case appears in a clinician worklist with basic metadata and a summary of the main findings.

Clinicians open individual cases to review both the underlying ECG and the corresponding AI-generated interpretations as shown in Figure 4. The main case view provides access to the raw traces, key interval measurements and an overview of detected rhythms and episodes. Users can navigate through the recording, adjust the display and examine segments that the system has highlighted as containing potential arrhythmias or other events of interest. The AI outputs are presented as suggestions that can be confirmed, modified or rejected by the clinician before finalisation.

After review, the clinician can generate a structured report that combines automatically derived measurements and rhythm classifications with their own edits and comments. Reports are stored in the platform for audit and follow-up and can be exported or interfaced with external electronic health record systems as required. At an organisational level, aggregate views of system usage and case mix (for example, number of analysed recordings and distribution of reported rhythms) support monitoring, quality assurance and service planning.

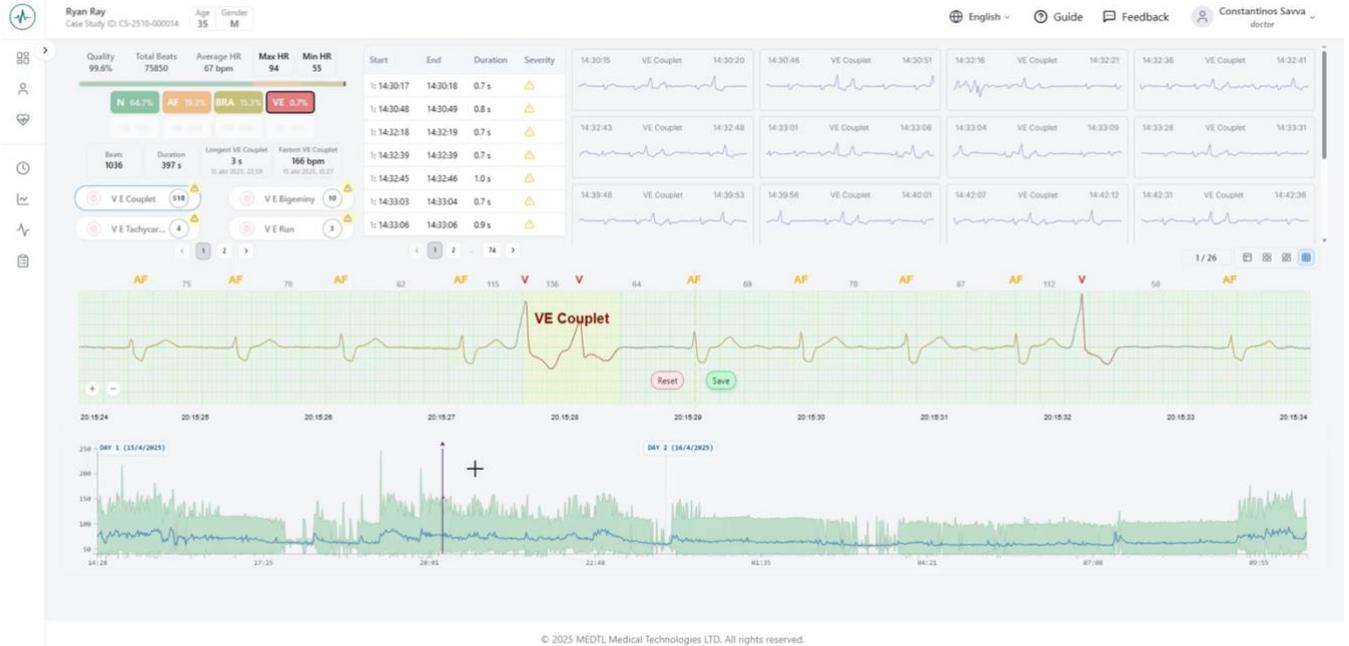

Fig. 4. Event-analysis view in the AI-HEART clinical user interface.

### 3.5. Development Process And Quality Management

The software development process for AI-HEART follows a documented lifecycle aligned with medical-device software standards. Requirements, design specifications, test cases and risk analyses are maintained in controlled repositories. Changes to the application and AI components are managed through version control and undergo peer review and automated testing before deployment to staging and production environments.

A quality-management system aligned with ISO 13485 provides the framework for managing documentation, design reviews, verification and validation activities, incident handling, and post-deployment monitoring. The platform architecture and tooling are chosen to support traceable links between requirements, implementation and test evidence, which is essential for future regulatory submissions.

## 4. Experimental Results

### 4.1. Training Performance

The overall training performance is summarized in Table I. GPU NVIDIA A100-SXM4-40GB was used for training the algorithms.

TABLE I
OVERALL TRAINING PERFORMANCE

| Algorithm | Number of model parameters (MB) | Number of epochs |
| --- | --- | --- |
| Delineation | 21.92 | 80 |
| Beat classification | 26.01 | 40 |
| Rhythm classification | 26.01 | 40 |

## 4.2. Delineation Performance

The delineation model was evaluated on the internal test set using the reference annotations described in Section 2.3. For each lead and annotated wave, predicted onset and offset positions were compared with ground truth at the sample level. Table II summarises the overall performance, reported as per-sample accuracy, tolerant accuracy within a fixed temporal window, F1-score and AUC.

Across all annotated segments, the model achieved a per-sample accuracy of 93.9%, with a tolerant accuracy of 98.4% within ±5 samples. The corresponding F1-score and AUC were 0.94 and 0.997, respectively (Table II). These results indicate that the model is able to locate P, QRS and T waves with an accuracy sufficient for automated interval measurement and for supporting downstream rhythm classification. Visual inspection of test cases confirmed that residual errors are typically confined to borderline regions between waves or to segments affected by substantial noise.

TABLE II
PERFORMANCE OF THE ECG DELINEATION MODEL ON THE INTERNAL TEST SET

| Metric | Value |
|---|---|
| Accuracy (95% CI), % | 93.9 (93.8-94.1) |
| Tolerant accuracy (%) (±5 samples) | 98.4 |
| Precision (95% CI), % | 93.8 |
| F1-score (95% CI) | 0.94 (0.99-0.99) |
| AUC | 0.997 |

## 4.3. Noise Detection

The noise-detection model was assessed on the test set using segment-level labels for "clean" and "noisy" recordings. Table III reports accuracy, precision, recall, F1-score and AUC. The model achieved an accuracy of 99.05%, precision of 99.20%, recall of 99.01% and an F1-score of 0.991, indicating that low-quality segments can be reliably identified and flagged for exclusion or manual review. In practice, this reduces the number of misclassifications attributable to obvious artefacts, such as electrode detachment, motion noise or severe baseline instability, and supports more stable arrhythmia analysis across heterogeneous recording conditions.

TABLE III
PERFORMANCE OF THE NOISE DETECTION MODEL

| Metric | Value |
|---|---|
| Accuracy (95% CI), % | 99.05 (98.98 - 99.11) |
| Precision (95% CI), % | 99.20 (99.11-99.27) |
| Recall (95% CI), % | 99.01 (95% CI 98.91-99.09) |
| F1-score (95% CI) | 0.991 (95% CI 0.9-99.09) |
| AUC | 0.997 |

## 4.4. Beat Classification

Beat classifier was first evaluated on a core set of rhythm classes comprising atrial fibrillation (AF), atrial flutter (AFL), normal sinus rhythm (N), junctional rhythm (J), supraventricular ectopic beats (S), ventricular ectopic beats (V) and an unspecified/noise class (X). Table IV presents per-class sensitivity, specificity, precision, F1-score and accuracy. Values are reported as % (95% CI) in Table IV.

For the beat level set, per-class sensitivities were generally in the high nineties, with specificities close to or at 100% for most classes. Overall accuracies for AF, AFL, N, J, S, V and X were approximately 99% or higher. Misclassifications were rare and typically involved confusion between supraventricular and ventricular ectopy in borderline examples or between noisy segments and the X class in low-quality recordings. Macro-averaged and micro-averaged F1-scores across the core classes were both close to 0.99, indicating balanced performance. In this subsection, AF, AFL, N, J, S, V and X refer to beat-level labels assigned to individual QRS complexes.

TABLE IV
BEAT CLASSIFICATION RESULTS

| Metric | AF (95% CI) | AFL (95% CI) | J (95% CI) | N (95% CI) | S (95% CI) | V (95% CI) | X (95% CI) |
|---|---|---|---|---|---|---|---|
| Sensitivity (Recall, %) | 98.7 (98.1–99.3) | 96.5 (96.0–97.0) | 99.4 (98.9–99.9) | 98.5 (97.8–99.2) | 98.4 (97.2–99.6) | 99.0 (98.1–99.9) | 97.2 (96.3–98.1) |
| Specificity (%) | 99.6 (99.3–99.9) | 99.9 (99.8–1) | 99.9 (99.8–1) | 99.6 (99.3–99.9) | 99.6 (99.3–99.9) | 99.8 (99.7–99.9) | 99.8 (99.7–99.9) |
| Precision (%) | 98.8 (98.1–99.5) | 97.0 (96.4–97.6) | 98.7 (98.0–99.4) | 99.0 (98.6–99.4) | 97.6 (97.0–98.2) | 98.9 (98.1–99.7) | 96.7 (96.0–97.4) |
| F1-score (%) | 98.7 (98.1–99.3) | 96.7 (96.1–97.3) | 99.0 (98.6–99.4) | 98.8 (98.2–99.4) | 98.0 (97.6–98.4) | 98.8 (98.3–99.3) | 97.1 (94.4–96.8) |
| Accuracy (%) | 99.3 (99.1–99.5) | 99.7 (99.6–99.8) | 99.9 (99.8–99.9) | 99.3 (99.2–99.4) | 99.4 (99.3–99.5) | 99.6 (99.4–99.8) | 99.6 (99.5–99.7) |

AF = atrial fibrillation; AFL = atrial flutter; N = normal rhythm; V = ventricular ectopic; S = supraventricular ectopic; J = junctional rhythm; X = unspecified/noise. Entries are metric % (95% CI).

### 4.5. Arrhythmia Classification

The rhythm-level classifier was also evaluated on a rhythm label set that includes first-, second- and third-degree atrioventricular block (AV1, AV21, AV22, AV3), sinus arrhythmia (SA) and supraventricular tachycardia (SVT), in addition to AF, AFL, normal sinus rhythm (N) and an unspecified/noise class (X). These classes are less frequent in the internal dataset, which presents a greater challenge for model training.

Despite class imbalance, the rhythm-level model achieved very high specificities (97–100%) across all classes in this rhythm label set, with sensitivities between 76.5% and 98.9% and F1-scores between 80.0% and 98.5% (Table V). Sinus arrhythmia and SVT, which were better represented in the dataset, showed F1-scores of 88.9% and 88.5% and accuracies of 99.9% and 99.8%, respectively, comparable to the core beat-level classes. For the AV block subtypes, sensitivities ranged from 76.5% for second-degree AV block type I (AV21) to 97.4% for first-degree AV block (AV1), with corresponding F1-scores from 80.0% to 93.3%, indicating reliable detection of both first-degree and higher-degree block despite their lower prevalence. Macro-averaged sensitivity and F1-score across this ten-class rhythm-level set were approximately 89–90%, with macro specificity above 99% and average accuracy around 99.5%.

TABLE V
ARRHYTHMIA CLASSIFICATION RESULTS

| Metric | AF (95% CI) | AFL (95% CI) | AV1 (95% CI) | AV21 (95% CI) | AV22 (95% CI) | AV3 (95% CI) | N (95% CI) | SA (95% CI) | SVT (95% CI) | X (95% CI) |
|---|---|---|---|---|---|---|---|---|---|---|
| Sensitivity (Recall, %) | 97.3 (97.1–97.5) | 95.0 (94.88–95.2) | 97.4 (97.2–97.6) | 76.5 (76.3–76.7) | 80.0 (79.88–80.2) | 83.3 (83.1–83.5) | 98.9 (98.7–99.1) | 85.7 (85.5–85.9) | 91.1 (90.9–91.3) | 89.9 (89.7–90.1) |
| Specificity (%) | 99.3 (99.1–99.5) | 99.6 (99.58–99.7) | 99.9 (99.8–1) | 99.9 (99.8–1) | 99.9 (99.8–1) | 99.8 (99.78–99.9) | 98.1 (98.0–98.2) | 99.9 (99.8–1) | 99.8 (99.78–99.9) | 99.7 (98.6–99.9) |

| Precision (%) | 98.1(98.0–98.2) | 94.7 (94.5–94.9) | 89.4 (89.2–89.6) | 87.8 (87.6–88.0) | 80.0 (79.98–80.1) | 80.0( 79.8–80.2) | 98.1 (98.0–98.2) | 92.3 (92.1–92.5) | 86.0 (85.8–86.2) | 94.7 (94.5–94.9) |
|---|---|---|---|---|---|---|---|---|---|---|
| F1-score (%) | 97.7 (97.5–97.9) | 94.9 (94.78–95.1) | 93.3 (93.1–93.5) | 81.8 (81.7–81.9) | 80.0 (79.9–80.1) | 80.0 (79.8–80.2) | 98.5 (98.4–98.6) | 88.9 (88.78–89.1) | 88.5 (88.4–88.6) | 92.3 (92.1–92.5) |
| Accuracy (%) | 98.8 (98.7–98.9) | 99.6 (99.58–99.7) | 99.9 (99.8–1) | 99.9 (99.8–1) | 99.9 (99.8–1) | 99.8 (99.78–99.9) | 98.1 (98.0–98.2) | 99.9 (99.8–1) | 99.8 (99.78–99.9) | 99.0 (98.9–99.1) |

AF = atrial fibrillation; AFL = atrial flutter; AV1 = first-degree AV block; AV21 = second-degree AV block type I; AV22 = second-degree AV block type II; AV3 = third-degree AV block; N = normal rhythm; SA = sinus arrhythmia; SVT = supraventricular tachycardia; X = unspecified/noise. Entries are metric % (95% CI)

### 4.6. Impact of Data Augmentation and Curation

To quantify the effect of GAN-based augmentation and expert-in-the-loop dataset curation, models trained with and without these techniques were compared on the same test set. For under-represented classes such as specific AV block subtypes and SVT, the use of synthetic examples and relabelled data led to consistent improvements in sensitivity and F1-score, typically in the range of several percentage points, while leaving performance on abundant classes essentially unchanged.

The Dataset Tornado curation tool leverages low-dimensional embeddings computed using Uniform Manifold Approximation and Projection (UMAP) to visualise the structure of the feature space. Embedding visualisations generated before and after curation showed that, following expert relabelling, clusters corresponding to each rhythm became more compact and better separated (Fig. 5). This pattern suggests that a significant portion of the original errors stemmed from labelling noise and ambiguous examples rather than model limitations. These observations support the inclusion of explainability and curation tools as part of the overall AI-development workflow and as enablers of continuous improvement in a regulated setting.

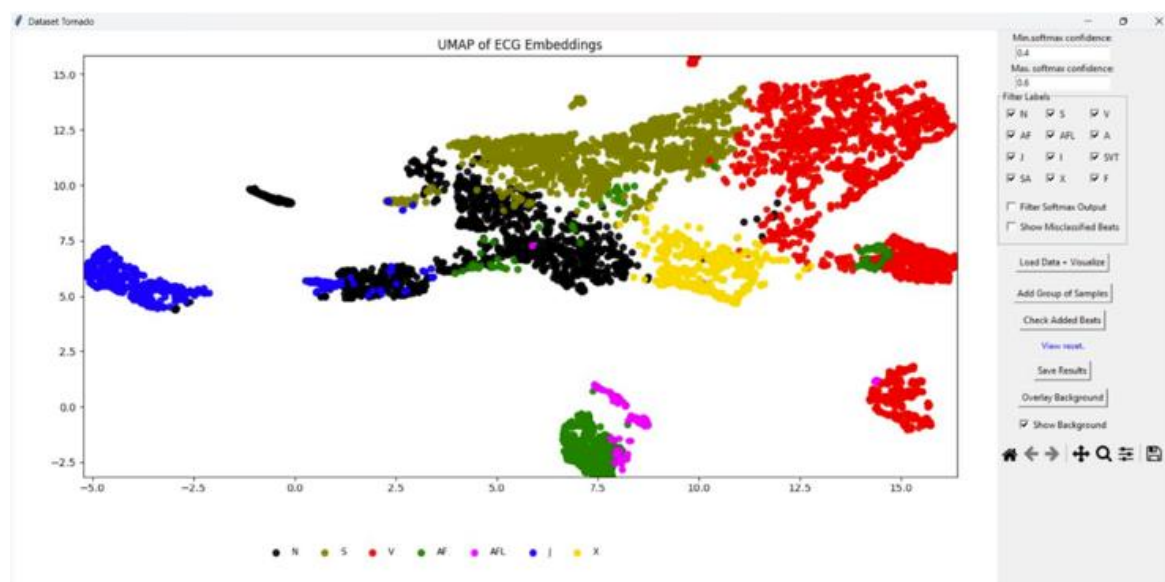

Fig. 5. UMAP of ECG embeddings generated by the Dataset Tornado tool, illustrating class-cluster structure and supporting expert-in-the-loop dataset curation.

### 4.7. Runtime and Scalability

Finally, the runtime performance of the AI pipeline was measured in a representative deployment configuration. For standard 10-second segments extracted from three-lead ambulatory recordings, end-to-end inference time, including preprocessing, delineation, noise detection and arrhythmia classification, was on the order of a fraction of a second per case on a commodity server-class CPU, and substantially lower when a GPU was available. Batch processing of larger volumes of recordings scales linearly with the number of instances assigned to the AI service.

These results indicate that the platform is capable of handling real-time or near-real-time analysis in typical clinical workflows and can be scaled horizontally in the cloud to accommodate higher workloads.

## 5. Discussion

### 5.1. Summary of Findings

This paper presents the design and internal evaluation of AI-HEART, a cloud-based AI-ECG platform that integrates distinct deep-learning models for ECG delineation, noise detection and multi-class classification at both beat and rhythm level. The delineation model achieved a per-sample accuracy of 94.3% and a tolerant accuracy of 98.4% within ±5 samples, with an F1-score of 0.94 and an AUC of 0.997 on the internal test set. The noise-detection model reached 99.05% accuracy, with precision, recall and F1-score all close to 0.99, indicating that low-quality segments can be reliably identified and flagged. For beat-level classification, the dedicated beat classifier evaluated on the core beat label set (AF, AFL, N, J, S, V, X) showed sensitivities and specificities predominantly in the high nineties, with per-class accuracies around 99% and macro- and micro-averaged F1-scores close to 0.99. For rhythm-level arrhythmia classification, a separate rhythm classifier evaluated on the ten-class rhythm label set that includes AF, AFL, multiple degrees of AV block, sinus arrhythmia, supraventricular tachycardia, normal sinus rhythm and a noise class achieved specificities at or close to 100%, with macro-averaged sensitivity and F1-score around 89–90%. AV block subtypes attained sensitivities of at least 76% and F1-scores between 80% and 93%, indicating clinically useful performance even for these comparatively rare rhythms. Together, these results indicate that AI-HEART can provide accurate, noise-aware ECG interpretation across a broad spectrum of rhythms, suitable for integration into clinical workflows.

### 5.2. Comparison With Existing AI-ECG Approaches

The reported performance is consistent with, and in several aspects complementary to, existing deep-learning approaches for ECG analysis described in the literature. Prior work has demonstrated that convolutional and recurrent neural networks can achieve high accuracy in arrhythmia detection, often exceeding traditional rule-based algorithms [5, 7]. Meta-analyses of AI-based atrial fibrillation detection from photoplethysmography and single-lead ECG signals have reported pooled sensitivities and specificities in the 90–96% and 95–97% range [16], which is broadly comparable to the AF performance observed in AI-HEART.

Commercial systems such as mobile ECG platforms and cloud-based holter analysis solutions have shown high performance for AF and other major arrhythmias in real-world cohorts [6, 12, 13]. However, many of these systems are optimised for specific device ecosystems or monitoring scenarios and often focus on a narrower set of conditions (for example, AF screening or Holter rhythm analysis). In contrast, this study is designed as a multi-class classifier covering both common rhythms and rarer conditions such as AV block subtypes and supraventricular tachycardia, with explicit mechanisms for noise handling and signal quality quantification. The combination of delineation, noise-aware preprocessing and broad rhythm coverage distinguishes the platform from AF-only or single-indication solutions and aligns with the need for more general ECG decision-support tools in routine practice.

The presence of a high-performing noise-detection model is particularly relevant in comparison with prior work, where the impact of signal quality is often managed implicitly through dataset curation or exclusion criteria. By making noise detection an explicit component of the pipeline, AI-HEART can provide transparent quality feedback and reduce misclassifications driven by artefacts, which is critical for deployment in primary care and resource-limited settings where acquisition conditions may be suboptimal.

### 5.3. Clinical Implications and Potential Use Cases

The results suggest several potential use cases for AI-HEART. In primary care or emergency settings, the platform could be used to provide rapid, automated pre-interpretation of multi-lead ECGs. In this study we evaluated three-lead Holter recordings; extension and validation on standard 12-lead ECGs represent an important future step. In hospitals or tele-cardiology services, AI-HEART could function as a decision-support tool, pre-screening ECGs and prioritising cases that require urgent expert attention. The ability to quantify signal quality and to clearly flag segments that are too noisy for reliable interpretation may help avoid false alarms and support more consistent reporting standards.

The multi-class nature of the classifier also supports longitudinal monitoring scenarios, where changes in rhythm patterns over time, such as the emergence of paroxysmal AF, progression of AV block or increasing ectopic burden, can be tracked systematically. Furthermore, the cloud-based architecture and configurable workflow make it possible to adapt the platform to different organisational contexts, from single-clinic deployments to multi-site networks.

It is important to emphasise that, in all these scenarios, AI-HEART is intended to augment rather than replace clinician judgement. The user interface is explicitly designed to allow clinicians to review, modify and override AI-generated interpretations, and to incorporate their own findings into the final report. This "human-in-the-loop" design is aligned with regulatory expectations for AI-based medical devices and with practical considerations for clinician acceptance.

### 5.4. Robustness, Generalisation and Continuous Learning

A recurring challenge for AI-ECG systems is robustness across devices, populations and recording conditions. The explicit modelling of noise in AI-HEART addresses one aspect of this challenge by allowing low-quality segments to be excluded or flagged, thereby reducing the risk of erroneous classifications driven by artefacts. The strong performance of the noise-detection model suggests that this approach is effective in the internal dataset and provides a basis for robust deployment in environments with variable acquisition quality.

Another challenge is the handling of rare arrhythmias and edge cases, where limited training data can lead to unstable performance. The incorporation of GAN-based augmentation and UMAP-based dataset curation is a pragmatic response to this problem. The observed improvements in sensitivity and F1-score for AV block subtypes and SVT after augmentation and relabelling indicate that, when carefully controlled, synthetic data and expert-in-the-loop curation can help mitigate class imbalance without degrading performance on more prevalent rhythms.

These tools also contribute to continuous learning. By enabling systematic identification and correction of labelling errors or ambiguous examples, the platform can evolve as more data are collected, while still operating under a quality-management framework compatible with medical-device regulations. In future post-market settings, similar mechanisms could support monitored model updates under a predetermined change control plan, consistent with emerging regulatory guidance.

### 5.5. Information Management Implications

AI-HEART is not only a set of deep-learning models but also an information system for managing long-duration ambulatory ECG data and derived annotations. The platform's modular, cloud-based architecture allows ECG recordings, intermediate model outputs and clinician edits to be stored under a unified data model with consistent identifiers and audit trails. This design supports traceability of each automated decision back to the original signal and associated model version, which is essential for quality management, post-market surveillance and regulatory audits. At the same time, separating the user-facing interface from back-end AI services makes it possible to update or scale individual components without disrupting clinical workflows.

From an information governance perspective, AI-HEART illustrates how regulatory and organisational requirements shape data flows in AI-enabled clinical systems. Requirements related to ISO 13485 and anticipated MDR/FDA certification influence decisions about logging granularity, retention of raw and processed data, and the documentation of training and validation datasets. Recent evidence from hospital management also highlights governance gaps around data quality, documentation/metadata, security, and ethical access, factors that directly shape how AI-enabled clinical data pipelines should be designed and operated [22]. Regional deployment and data-residency constraints further affect where recordings and reports are stored and how cross-border access is managed. The system's human-in-the-loop workflow, in which clinicians review, correct and approve AI-generated annotations before final reporting, also embeds feedback into the information lifecycle: clinician edits can be captured as structured data for future model refinement, while ensuring that legal responsibility for the diagnostic interpretation remains with the human user.

For healthcare organisations, these design choices have practical implications. Integrating AI-HEART with existing ECG management systems or electronic health records requires alignment of identifiers, consent and access-control policies, rather than merely deploying a standalone algorithm. More broadly, the case suggests that successful deployment of AI-ECG services depends as much on robust information management, covering data capture, storage, provenance, access and auditability, as on model accuracy. Platforms that explicitly incorporate these considerations are better positioned to support sustainable, large-scale use of AI in routine clinical practice.

### 5.6. Limitations And Future Work

The current evaluation has several limitations. First, all results are based on internal datasets and internal test splits. Although patient-level partitioning reduces the risk of information leakage, external validation on independent cohorts from different institutions and populations will be necessary to confirm the generalisability of the models. Second, the dataset used for rare arrhythmias remains comparatively small despite augmentation, and some performance metrics for the rarest classes have wide uncertainty; prospective multicentre data collection will be required to further stabilise these estimates.

Third, the present study focuses on rhythm analysis and signal quality; structural markers such as left ventricular dysfunction or other ECG-derived risk scores, which have been explored in other AI-ECG work [10], are not yet integrated into the deployed model set. Extending AI-HEART to incorporate such markers is a natural next step and could enhance its utility for screening and risk stratification. Finally, while the system architecture and user workflow have been designed with regulatory and clinical requirements in mind, this paper does not report formal usability testing or prospective clinical impact studies, which will be critical to demonstrate real-world benefit.

Future work will therefore focus on external validation in diverse clinical environments, including under-served settings; prospective studies assessing workflow impact, diagnostic yield and clinician trust; and the integration of additional predictive models for structural and functional cardiac abnormalities. Further development of post-deployment monitoring and change-management processes will also be essential to support safe model evolution in line with regulatory expectations.

**CRediT authorship contribution statement**

**Artemis Kontou:** Writing – original draft, Project administration, Conceptualization. **Natalia Miroshnikova:** Writing – review & editing, Data curation, Methodology, Formal analysis, **Costakis Matheou:** Writing – review & editing, Project administration, Resources. **Sophocles Sophocleous:** Methodology, Formal analysis, Data curation. **Nicholas Tsekouras:** Software. **Kleanthis Malialis:** Writing – review & editing. **Panayiotis Kolios:** Writing – review & editing.

**Declaration of Competing Interest**

Several authors are employed by MEDTL Medical Technologies Ltd., the developer and owner of the AI-HEART platform evaluated in this study. The ECG datasets used for model development and evaluation are held by MEDTL Medical Technologies. These relationships are disclosed to ensure transparency. The authors declare no other competing financial interests or personal relationships that could have influenced the work reported in this paper.

**Data Availability**

The ECG datasets used to develop and evaluate AI-HEART consist of de-identified long-duration three-lead ambulatory electrocardiogram recordings collected in routine clinical care and internal annotated repositories held by MEDTL Medical Technologies. Due to patient privacy considerations and contractual constraints with data providers, these datasets cannot be made publicly available. Aggregated performance metrics and model-evaluation summaries are available from the corresponding author upon reasonable request.

**Declaration of Generative AI Use**

No Generative AI was used to prepare this study.